\documentclass[9pt, conference]{IEEEtran}
\IEEEoverridecommandlockouts
\usepackage{cite}
\usepackage{amsmath,amssymb,amsfonts}
\usepackage{algorithmic}
\usepackage{graphicx}
\usepackage{textcomp}
\usepackage{xcolor}
\usepackage{multirow}
\usepackage{array}
\usepackage{booktabs}
\usepackage{float}
\usepackage{graphicx}
\usepackage{subcaption}
\usepackage{url}
\def\BibTeX{{\rm B\kern-.05em{\sc i\kern-.025em b}\kern-.08em
    T\kern-.1667em\lower.7ex\hbox{E}\kern-.125emX}}
\begin{document}
\setlength{\floatsep}{2pt}
\setlength{\textfloatsep}{5pt}
\setlength{\intextsep}{2pt}
\title{Efficient and Effective Model Extraction\\
}

\author{
    \Large 
    \textit{Hongyu Zhu\textsuperscript{1}, Wentao Hu\textsuperscript{1}, Sichu Liang\textsuperscript{2}, Fangqi Li\textsuperscript{1}, Wenwen Wang\textsuperscript{3}, Shilin Wang\textsuperscript{1}} 
    \vspace{0.3cm} 
    \\
    \textsuperscript{1}School of Electronic Information and Electrical Engineering, Shanghai Jiao Tong University, China\\
    \textsuperscript{2}School of Artificial Intelligence, Southeast University, China \hspace{0.3cm} 
    \textsuperscript{3}Thorough Future, China
}

\maketitle

\begin{abstract}
Model extraction aims to steal a functionally similar copy from a machine learning as a service (MLaaS) API with minimal overhead, typically for illicit profit or as a precursor to further attacks, posing a significant threat to the MLaaS ecosystem. However, recent studies have shown that model extraction is highly inefficient, particularly when the target task distribution is unavailable. In such cases, even substantially increasing the attack budget fails to produce a sufficiently similar replica, reducing the adversary’s motivation to pursue extraction attacks. In this paper, we revisit the elementary design choices throughout the extraction lifecycle. We propose an embarrassingly simple yet dramatically effective algorithm, \underline{E}fficient and \underline{E}ffective Model \underline{E}xtraction (\(E^3\)), focusing on both query preparation and training routine. \(E^3\) achieves superior generalization compared to state-of-the-art methods while minimizing computational costs. For instance, with only $0.005\times$ the query budget and less than $0.2\times$ the runtime, \(E^3\) outperforms classical generative model based data-free model extraction by an absolute accuracy improvement of over 50\% on CIFAR-10. Our findings underscore the persistent threat posed by model extraction and suggest that it could serve as a valuable benchmarking algorithm for future security evaluations.
\end{abstract}


\section{Introduction}
Machine Learning as a Service (MLaaS) APIs from major companies like Microsoft, Google, and OpenAI provide users with access to powerful deep learning models via a pay-per-query interface. These models span applications such as visual recognition, natural language processing, speech analysis, and code generation, benefiting both industry and everyday life 
\cite{a1}\ - \cite{codex}
. However, the public availability of these APIs raises concerns about model theft \cite{a4}. Besides direct illegal distribution of model parameters \cite{a5} - \cite{a7}, clients may attempt to reverse-engineer the models through \textbf{model extraction}, a tactic that is gaining increasing attention \cite{a8}.

In model extraction, an adversary queries a victim API and uses the resulting probability vectors to train a surrogate model that approximates the victim’s functionality \cite{a8}\cite{T. Orekondy}. In data-dependent scenarios, the adversary samples in-distribution (IND) queries, while in data-free scenarios, they lack prior knowledge of the target distribution and synthesize substitute queries to reveal the private knowledge. The process operates in a black-box setting, where the adversary only has access to query inputs and responses, without any information about the victim’s architecture, parameters, or gradients. It is assumed that model extraction can achieve comparable performance with minimal effort, bypassing the costs of data labeling and training from scratch. This poses a significant threat to the intellectual property and confidentiality of models 
\cite{T. Orekondy} - \cite{a9}
, making model extraction one of the most pressing concerns in industrial machine learning \cite{a10}.

However, the implicit assumption that model extraction is cheaper in terms of data or computation \cite{T. Orekondy}\cite{a11} has recently been questioned \cite{A. Shafran}. Building a model primarily involves data preparation, computational resources, and expertise. In data-dependent scenarios, adversaries must still gather IND samples, and using the victim as a labeling oracle may not significantly reduce costs compared to established data pipelines \cite{A. Shafran}. In data-free extraction, while the need for IND samples is avoided, adversaries often rely on computationally intensive generative models to synthesize samples \cite{J. B. Truong}\cite{J. Zhang}, which can demand significantly more queries (e.g., 20 million) and longer training cycles to converge \cite{J. B. Truong}\cite{A. Shafran}. Moreover, implementing sophisticated extraction algorithms often requires more expertise than training a model from scratch \cite{J. Zhang}\cite{P. Karmakar}. Thus, achieving a high-performing model with minimal resources is challenging, and the cost-effectiveness of model extraction has been exaggerated.

\vspace{-5pt}
\begin{figure}[htbp]
\centerline{\includegraphics[width=0.9\linewidth]{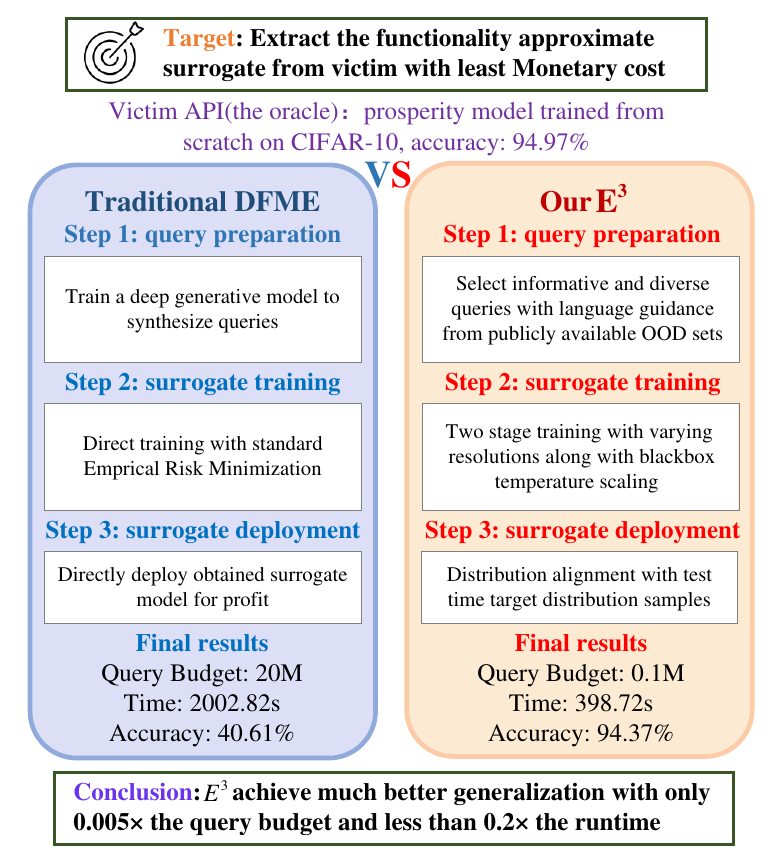}}  
\caption{Comparison of $E^3$ with traditional DFME.}
\label{fig:E3}
\end{figure}
\vspace{-5pt}

Does this mean model extraction is entirely ineffective? In this paper, we revisit key design choices throughout the lifecycle of model extraction and reveal the inefficiency of mainstream methods. We further propose surprisingly simple techniques that achieve impressive generalization with minimal cost. For \textbf{query preparation}, we show that sophisticated deep generative models are unnecessary when IND samples are unavailable. Instead, publicly available OOD samples, selected via language-guidance, outperform synthetic samples in functionality approximation. For  \textbf{surrogate training}, low-resolution pretraining significantly reduces costs while enhancing victim guidance. We also mathematically prove, for the first time, that temperature scaling—previously overlooked in the extraction community—is applicable in black-box settings and accelerates convergence while improving generalization. Finally, during \textbf{deployment stage}, we introduce a novel perspective where unlabeled target distribution samples naturally encountered by the surrogate can be leveraged for rapid unsupervised adaptation, bridging the distribution gap between training and testing in data-free scenarios.

By incorporating these nearly cost-free components, we propose a novel $E^3$ extraction algorithm that achieves state-of-the-art performance with the lowest query budget and computational cost compared to previous methods. This serves as a proof of concept that model extraction continues to pose a significant threat to MLaaS. 
Figure \ref{fig:E3} highlights the effectiveness of our approach.

Our contributions are summarized as follows:
\begin{itemize}
\item We highlight the inefficiency of current model extraction techniques, identifying resource-intensive yet ineffective components by revisiting the extraction pipeline.
\item At each extraction stage, we introduce computationally efficient strategies that optimize both efficiency and effectiveness under a fixed query budget, resulting in the $E^3$ algorithm.
\item With the lowest query budget and runtime, we achieve significant improvements in generalization, confirming that model extraction remains a substantial threat.
\end{itemize}

\section{Related Work}

In model extraction, the adversary \( \mathcal{A} \) prepares a query set \( D^Q \) to access a \textit{black-box} victim API \( f_{\theta_v} \), obtaining response probability vectors \( y = f_{\theta_v}(x) \) for each sample \( x \in D^Q \). Early \textit{data-dependent} extraction strategies\cite{T. Orekondy} select IND samples as \( D^Q \), sometimes assuming direct access to the victim training set \cite{H. Zhu}. A surrogate model \( f_{\theta_s} \) is then trained via knowledge distillation \cite{G. Hinton} to replicate \( f_{\theta_v} \). This process is formalized as:
\newenvironment{shrinkeq}[1]
{ \bgroup
\addtolength\abovedisplayshortskip{#1}
\addtolength\abovedisplayskip{#1}
\addtolength\belowdisplayshortskip{#1}
\addtolength\belowdisplayskip{#1}}
{\egroup\ignorespacesafterend}
\begin{shrinkeq}{-1ex}
\begin{equation}
\theta_s^* = \arg \min_{\theta_s} KLD(f_{\theta_v}(x) || f_{\theta_s}(x))
\footnote{In model extraction, to reduce labeling costs, cross-entropy between predictions and ground-truth labels is typically omitted from the loss function.}.
\end{equation}
\end{shrinkeq}

Here, \( KLD(\cdot || \cdot) \) represents the Kullback-Leibler divergence, measuring the difference between the prediction distributions of \( f_{\theta_s} \) and \( f_{\theta_v} \). Acquiring IND sample, even without labels, is costly and challenging in privacy-sensitive contexts. Consequently, \textit{data-free} extraction, which does not rely on target distribution samples, has garnered widespread attention. Inspired by data-free knowledge distillation \cite{G. Fang}, synthesizing queries with \textbf{deep generative models} has become mainstream \cite{J. B. Truong}\cite{J. Zhang}. DFME \cite{J. B. Truong} formulates a three-player game between the victim, surrogate and generator using a generative adversarial network framework \cite{I. Goodfellow}, with zero-order optimization to estimate gradients from the black-box \( f_{\theta_v} \). IDEAL \cite{J. Zhang} optimizes the generator based on the prediction entropy of \( f_{\theta_s} \), bypassing gradient estimation on \( f_{\theta_v} \) and significantly lowering the query budget. However, inverting training samples from the model is challenging, and the synthesized samples often deviate from the original distribution, lacking realistic visual patterns, which hinders the generalization of \( f_{\theta_s} \). Moreover, training the generator from scratch also introduces considerable computational overhead.

Another approach involves randomly sampling OOD samples and adding adversarial perturbations \cite{C. Szegedy} to maximize the prediction entropy of \( f_{\theta_v} \) \cite{H. Yu} \cite{Y. Zhao}. These high-entropy \textbf{adversarial samples} better characterize the decision boundary of \( f_{\theta_v} \), increasing information leakage from OOD queries. However, generating adversarial perturbations is time-consuming, and the responses are often erroneous or ambiguous, limiting \( f_{\theta_s} \) from learning well-generalized mappings.

Lastly, \textbf{active sampling}-based extraction progressively select representative samples from candidate OOD set as \( D^Q \) \cite{P. Karmakar} \cite{P. Ren} \cite{V. Chandrasekaran}. However, selecting samples with a limited number of queries to \( f_{\theta_v}\) is challenging, as the adversary must construct structural representation of the entire candidate set and recalculate selection criteria at each iteration. Notably, model extraction has also been adapted for legitimate purposes like model compression as a service \cite{Y. Xu} \cite{J. Hong}. Whether for legitimate or illicit use, designing cost-effective extraction strategies remains crucial.

\section{Methodology}
In this section we elaborate on our proposed strategies in $E^3$. The first two apply to both data-dependent and data-free scenarios, while the last two are specifically tailored for data-free model extraction.
\subsection{Two-Stage Extraction with Varying Resolution}\label{AA}
The ultimate goal of model extraction is to replicate the victim model's functionality at a lower cost than training from scratch. However, without ground-truth supervision, the surrogate often requires the same sample volume as the victim's training set and longer training times to converge \cite{A. Shafran}. To address this, we propose a two-stage training routine with varying resolutions (VarRes), balancing training cost and generalization.

In image classification, the victim typically processes square images of size \( R \times R \). When the query set is limited, using the same input for the surrogate can lead to overfitting. Therefore, in the first stage, we randomly crop a rectangle from the image, rescale it to \( r \times r \) (where \textbf{r \textless\ R}), and optimize the following loss function:
\newenvironment{shrinkeq2}[1]
{ \bgroup
\addtolength\abovedisplayskip{#1}
\addtolength\belowdisplayshortskip{#1}
\addtolength\belowdisplayskip{#1}}
{\egroup\ignorespacesafterend}
\begin{shrinkeq2}{-1ex}
\begin{equation}
\theta_s^* = \arg \min_{\theta_s} \text{KL}(f_{\theta_v}(x) \,||\, f_{\theta_s}(\text{transform}(x)))
\end{equation}
\end{shrinkeq2}

where \(\textit{transform}\) refers to the input transformation applied to \(f_{\theta_s}\), such as the \textit{RandomResizedCrop} operation \cite{b18} in PyTorch \cite{b19}. In the second stage, we fine-tune the surrogate at full resolution during the final epochs as in Equation 1. Notably, the victim always receives the original image \(x\) as input, ensuring that the query budget remains unchanged. Low-resolution training in the first stage reduces computational costs while encouraging robustness to input scale variations. As shown in Fig.~\ref{fig:model_epoch}, training the surrogate with the same input size as the victim causes the training loss to rapidly converge to zero, limiting further guidance from the victim. In contrast, VarRes preserves a meaningful prediction difference between the victim and the surrogate, enabling the surrogate to continuously absorb task-specific knowledge from the victim. This teacher-student discrepancy is widely regarded as highly advantageous \cite{T. Orekondy}\cite{J. B. Truong}\cite{P. Karmakar}.


\noindent 
\begin{minipage}{0.49\columnwidth}
  \centering
  \includegraphics[width=\linewidth]{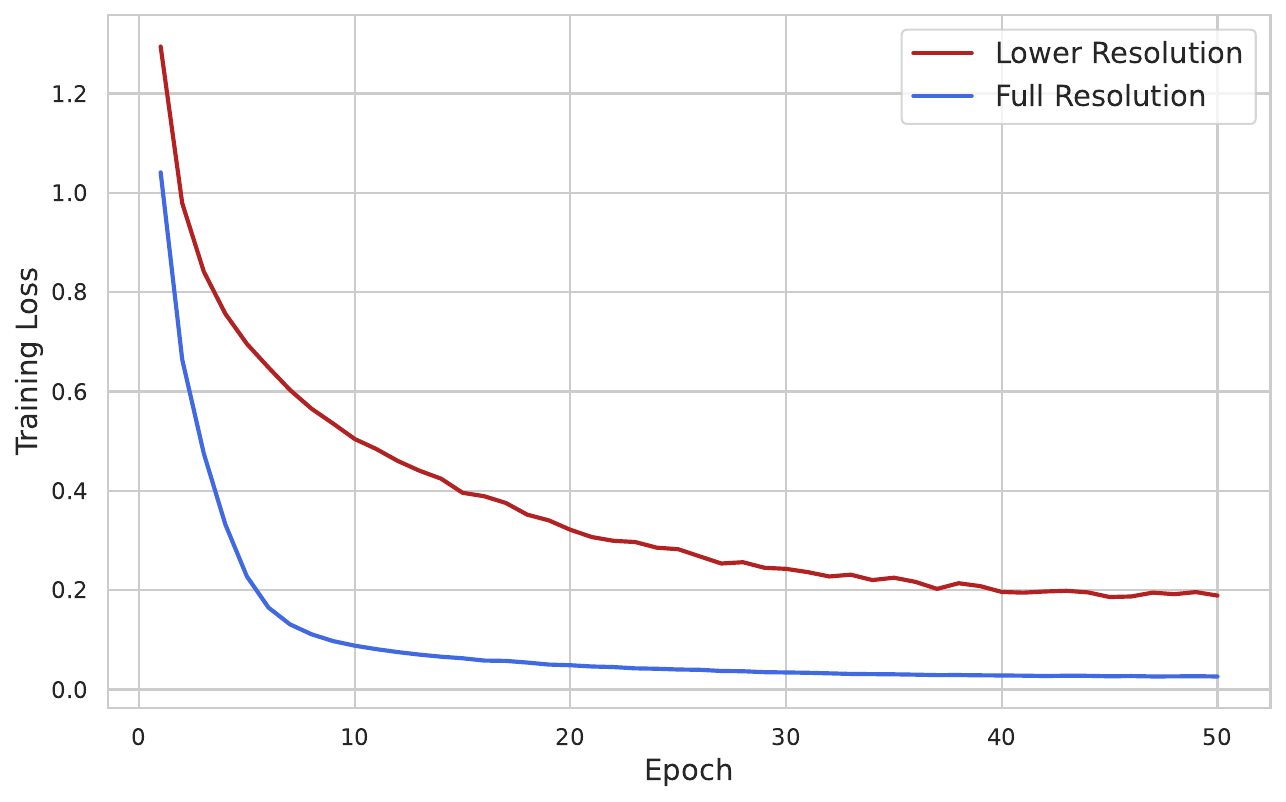} 
  \captionof{figure}{Training Loss at Low vs Full Resolution Across Epochs.} 
  \label{fig:model_epoch}
\end{minipage}
\hfill 
\begin{minipage}{0.49\columnwidth}
  \includegraphics[width=\linewidth]{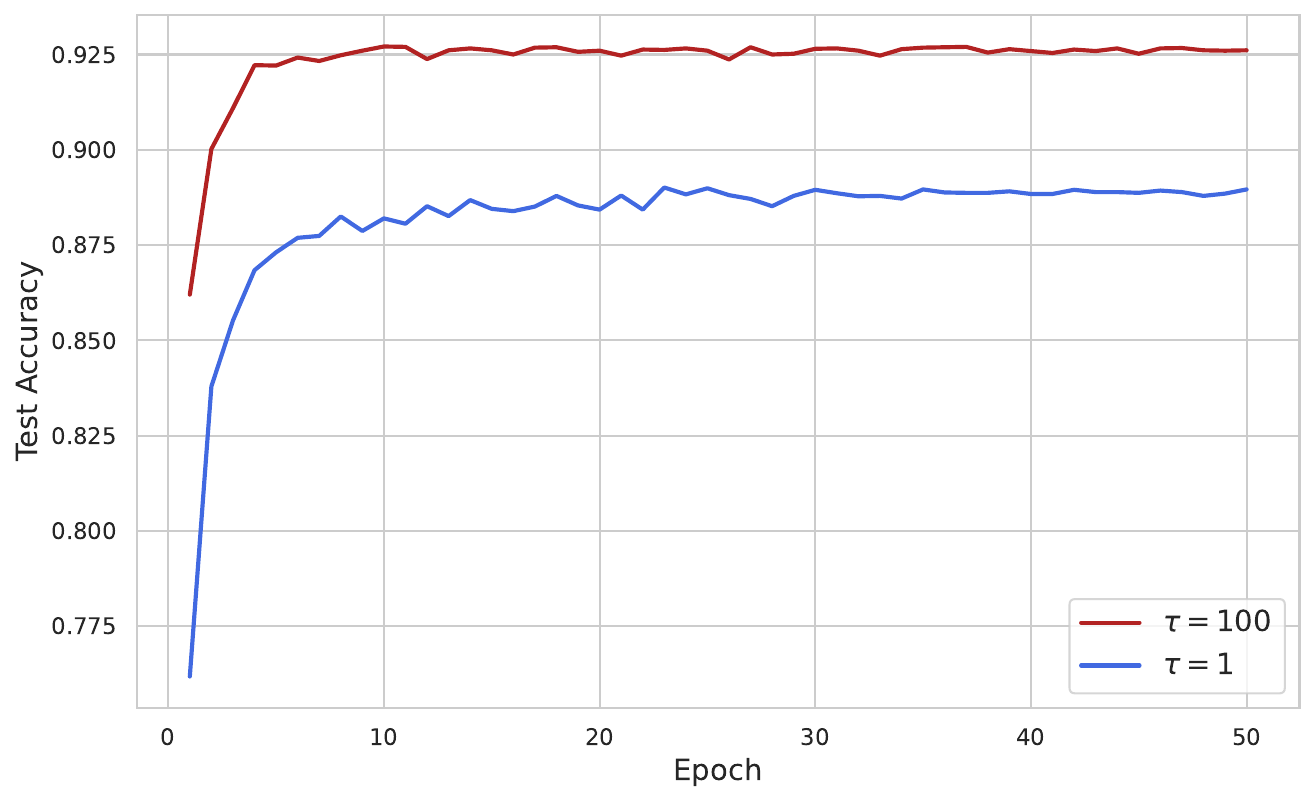} 
  \captionof{figure}{Test Accuracy with \(\tau\) = 1 or 100 Across Epochs.} 
  \label{fig:acc}
\end{minipage}%

\subsection{Temperature Scaling in Black-box Extraction}
When using KL divergence to measure the prediction difference between the victim and the surrogate, the optimization objective can be expressed as:
\newenvironment{shrinkeq3}[1]
{ \bgroup
\addtolength\abovedisplayskip{#1}
\addtolength\belowdisplayshortskip{#1}
\addtolength\belowdisplayskip{#1}}
{\egroup\ignorespacesafterend}
\begin{shrinkeq3}{-1ex}
\begin{equation}
\min -\sum_{j=1}^{c} \sigma_j(z_v) \log \sigma_j(z_s)
\end{equation}
\end{shrinkeq3}

where \(z_v\) and \(z_s\) are the output logits of \(f_{\theta_v}\) and \(f_{\theta_s}\), respectively, and are transformed into probability distributions via the softmax function \(\sigma(\cdot)\). However, the output of modern deep neural networks tends to be over-confident \cite{b20}, where only the predicted class \(t\) of the victim receives significant weight in \(\sigma_t(z_v)\), ignoring relative differences among other classes \cite{b21}. 
This results in near one-hot labels, making it difficult to capture the "dark knowledge" needed for the surrogate to effectively learn the victim's unique functionality.

To address this, the classic solution in knowledge distillation \cite{G. Hinton} is to soften the model's predictions with temperature scaling, transforming the prediction probabilities into \(\sigma(z_v / \tau)\) and \(\sigma(z_s / \tau)\), where \(\tau\) is the temperature parameter. This reduces class differences in the probabilities, allowing the loss function to focus on matching the entire logits vector \cite{b21}\cite{b22}. However, in model extraction, the adversary only has access to the black-box probability output \(\sigma(z_v)\) from the victim. Since \(\sigma(\cdot)\) is not injective and therefore non-invertible \cite{b23}, it was considered impossible to derive the specific temperature-scaled \(\sigma(z_v / \tau)\). 
As a result, temperature scaling has been overlooked in model extraction \cite{T. Orekondy} - \cite{P. Karmakar}.

In this paper, we show that simple elementary operations can derive \(\sigma(z_v / \tau)\) from \(\sigma(z_v)\) in a mathematically equivalent way.
\textbf{Proposition: }$\forall z \in {R^c},\tau  \in R,\sigma ({z_i}/\tau ) = \sigma (\log (\sigma({z_i}))/\tau )$\\
\textbf{Proof Sketch: }$\log (\sigma({z_i}))/\tau  = ({z_i} - \log (\sum\nolimits_j {\exp ({z_j})} ))/\tau = {z_i}/\tau  + C,$ \text{where} $ C = \log (\sum\nolimits_j {\exp ({z_j})}) /\tau. $
Since $\sigma(x)=\sigma(x+C)$\cite{H. Yu}, $\sigma(\log (\sigma({z_i}))/\tau)=\sigma({z_i}/\tau+C) =\sigma ({z_i}/\tau )$

This is the first demonstration that temperature scaling can be applied to black-box model extraction seamlessly. Fig.~\ref{fig:acc} shows a performance comparison of the surrogate on CIFAR-10, with and without temperature. Higher temperature significantly accelerates convergence and improves final generalization.

\subsection{Language-Guided Query Selection}
In data-free scenarios, the absence of IND samples for the target task makes it difficult to effectively transfer knowledge from the victim to the surrogate. Traditional approaches tackle this by synthesizing substitute samples using generative models or adversarial perturbations, both of which are computationally expensive. In this paper, we find that publicly available OOD samples from the Internet can offer valuable image priors, such as general shape and texture features. While the class semantics may not fully align with the target task, these samples are a cost-effective choice for queries. However, OOD samples that deviate significantly from the target distribution are less effective for knowledge transfer \cite{b24}\cite{b25}. The challenge, then, is efficiently constructing an optimal query set from these OOD samples. Active sampling \cite{P. Karmakar} progressively select the most informative samples from the OOD set as queries, but significantly increases the computational cost and query budget, as it requires generating representations or confidence scores for all samples using either the victim or surrogate \cite{P. Karmakar}\cite{b25}.

In this paper, we propose a novel multimodal strategy that leverages a language model to select a query set from an OOD set with a predefined class structure, semantically similar to the target task. Specifically, given a labeled OOD set \( D_O \) with class names \(\{C_{O_i}\}_{i=1}^{k_O}\) and target task class names \(\{C_{T_j}\}_{j=1}^{k_T}\) (which may not overlap), we compute the semantic similarity between each OOD class and the target task using a pretrained language encoder \( E_l(\cdot) \) as follows:
\newenvironment{shrinkeq4}[1]
{ \bgroup
\addtolength\abovedisplayskip{#1}
\addtolength\belowdisplayshortskip{#1}
\addtolength\belowdisplayskip{#1}}
{\egroup\ignorespacesafterend}
\begin{shrinkeq4}{-1ex}
\[
\text{similarity}_i =\sum_j \frac{ \left( E_l(C_{O_i}) \cdot E_l(C_{T_j}) \right)}{\left( |E_l(C_{O_i})| \cdot |E_l(C_{T_j})| \right)}
\]
\[
\text{similarity}_i' = \frac{\text{similarity}_i - \min(\text{similarity}_i)}{\max(\text{similarity}_i) - \min(\text{similarity}_i)}
\]
\end{shrinkeq4}

Thus, \(\text{similarity}_i' \times \left(\frac{|D^Q|}{|D_O|}\right)\) represents the sampling probability for each class in \(D_O\). This language-guided sampling strategy is extremely efficient, performed in a single pass without requiring additional queries to the victim. Moreover, it is model-agnostic, avoiding the bias towards either the victim or the surrogate that often arises in active sampling strategies.

\subsection{Test-Time Distribution Alignment}

\begin{table*}[tp]
\centering
\renewcommand{\arraystretch}{1.3} 
\caption{\small Comparison of $E^3$ with SOTA model extraction algorithms. AC: Adversarial Capability; DD: Data Dependent; DF: Data-Free. $Q_{\text{b}}$: Query Budget. $\uparrow$ (higher is better), $\downarrow$ (lower is better). Best results in \textbf{bold}, second-best in \textbullet.}
\resizebox{\textwidth}{!}{ 
\begin{tabular}{cccccccccccc}
\hline
\multirow[c]{4}{*}{\centering AC} & \multirow[c]{4}{*}{\centering Method} & \multicolumn{5}{c}{CIFAR-10} & \multicolumn{5}{c}{CIFAR-100} \\ \cline{3-12} 
&  & \multirow{2}{*}{\begin{tabular}[c]{@{}c@{}}$Q_{\text{b}}(\downarrow)$\end{tabular}} & \multicolumn{2}{c}{\begin{tabular}[c]{@{}c@{}}Victim: Resnet18\\ Acc: 94.97\%\end{tabular}} & \multicolumn{2}{c}{\begin{tabular}[c]{@{}c@{}}Victim: Resnet34 \\ Acc: 95.94\%\end{tabular}} & \multirow{2}{*}{\begin{tabular}[c]{@{}c@{}}$Q_{\text{b}}(\downarrow)$\end{tabular}} & \multicolumn{2}{c}{\begin{tabular}[c]{@{}c@{}}Victim: Resnet18\\ Acc: 77.33\%\end{tabular}} & \multicolumn{2}{c}{\begin{tabular}[c]{@{}c@{}}Victim: Resnet34\\ Acc: 81.88\%\end{tabular}} \\ \cline{4-7} \cline{9-12}

&  &  & Time($\downarrow$) & Acc($\uparrow$) & Time($\downarrow$) & Acc($\uparrow$) &  & Time($\downarrow$) & Acc($\uparrow$) & Time($\downarrow$) & Acc($\uparrow$) \\ \hline
\multirow[c]{3}{*}{\begin{tabular}[c]{@{}c@{}}DD\end{tabular}} & KnockoffNets & 0.05M & 395.28\textbullet{} & 91.56 ± 0.38 & 416.9\textbullet{} & 91.28 ± 0.86 & 0.05M & 395.2\textbullet{} & 71.65 ± 0.26 & 417.13\textbullet{} & 70.97 ± 1.29 \\
 & $E^3$ w/o VarRes & 0.05M & 394.93\textbullet{} & 94.11 ± 0.20\textbullet{} & 417.08\textbullet{} & 94.48 ± 0.37\textbullet{} & 0.05M & 395.19\textbullet{} & 75.95 ± 0.42\textbullet{} & 417.19\textbullet{} & 75.72 ± 0.21\textbullet{} \\
 & $E^3$ & \textbf{0.05M} & \textbf{343.58} & \textbf{95.48 ± 0.25} & \textbf{366.05} & \textbf{95.62 ± 0.14} & \textbf{0.05M} & \textbf{342.56} & \textbf{78.37 ± 0.43} & \textbf{364.43} & \textbf{79.16 ± 0.47} \\ \hline
\multirow[c]{7}{*}{\begin{tabular}[c]{@{}c@{}}DF\end{tabular}} & DFME & 20M & 2002.82 & 40.61 ± 2.39 & 2093.25 & 31.18 ± 2.68 & 20M & 2060.73 & 7.27 ± 0.92 & 2117.98 & 5.29 ± 1.34 \\
 & IDEAL & 0.25M & 4049.14 & 65.22 ± 1.88 & 4129.13 & 62.26 ± 1.68 & 0.3M & 4564.98 & 19.13 ± 0.58 & 4716.18 & 17.10 ± 0.40 \\
 & CloudLeak & 0.5M & 1860.66 & 87.65 ± 0.59 & 1894.72 & 85.48 ± 2.13 & 0.5M & 1867.77 & 53.89 ± 1.69 & 1884.91 & 48.82 ± 1.16 \\
 & SPSG & 5M & 1904.18 & 88.07 ± 0.42 & 1917.48 & 85.47 ± 0.49 & 5M & 1909.07 & 62.37 ± 0.76 & 1925.53 & 57.80 ± 1.21 \\
 & Marich & 0.1M & 958.75 & 90.66 ± 0.41 & 1029.60 & 88.85 ± 0.99 & 0.1M & 1171.92 & 69.49 ± 1.49 & 1265.86 & 69.98 ± 1.51 \\
 & $E^3$ w/o VarRes & 0.1M & 459.85\textbullet{} & 94.23 ± 0.17\textbullet{} & 482.44\textbullet{} & 93.22 ± 0.69\textbullet{} & 0.1M & 464.27\textbullet{} & \textbf{74.94 ± 0.56} & 485.72\textbullet{} & \textbf{72.71 ± 1.32} \\
 & $E^3$ & \textbf{0.1M} & \textbf{398.72} & \textbf{94.37 ± 0.26} & \textbf{421.39} & \textbf{94.01 ± 0.07} & \textbf{0.1M} & \textbf{401.86} & 72.31 ± 0.65\textbullet{} & \textbf{423.90} & 72.34 ± 1.73\textbullet{} \\ \hline
\end{tabular}
} 
\label{tab:comparison}
\end{table*}

Model extraction aims for the surrogate to replicate the behavior of the victim during deployment, ensuring that for a test sample \(x_{test}\), the surrogate’s output approximates the victim's conditional distribution, \( p_v(y|x_{test}) \). Although the surrogate is trained on \( p_v(y|x_{train}) \), which aligns with the desired behavior \( p_v(y|x_{test}) \) during testing, the absence of IND samples from the target task causes a mismatch. The marginal distribution of the surrogate’s training data \( p(x_{train}) \) inevitably differs from the deployment distribution \( p(x_{test}) \), leading to covariate shift \cite{b26} and impairing generalization. 
Recent data-free knowledge transfer methods alleviate this issue through distribution-invariant learning \cite{b27}, but they increase computational costs and require access to the victim’s internal features, which is unfeasible in black-box extraction.

Inspired by test-time classifier adjustment \cite{b28}, we propose a low-cost solution for unsupervised online adaptation, leveraging the unlabelled target distribution samples that the surrogate naturally encounters post-deployment, which enables the surrogate to better align with the target distribution. We decompose \(f_{\theta_s}\) into a feature extractor \(f_e(\cdot)\) and a classification head \(f_c(\cdot)\), where the classification head is parameterized by weights \(w \in \mathbb{R}^{c \times d}\) and biases \(b \in \mathbb{R}^c\), with \(c\) representing the number of classes and \(d\) the feature dimension output by \(f_e(\cdot)\). 
In prototype learning \cite{b29}, each \(w_i \in \mathbb{R}^d\) represents the prototype for class \(i\). The classification process of \(f_c(\cdot)\) can thus be interpreted as assigning a sample to the class with the highest similarity to its prototypes. Thus, we propose fine-tuning the classification head by updating the prototypes. For each class \(i\), we dynamically maintain a support set \(S_i = \{x_{i_j}\}_{j=1}^k\) consisting of the \(k\) samples with the lowest prediction entropy that \(f_{\theta_s}\) predicted as class \(i\). The prototype \(w_i\) is updated as follows:
\newenvironment{shrinkeq5}[1]
{ \bgroup
\addtolength\abovedisplayskip{#1}
\addtolength\belowdisplayshortskip{#1}
\addtolength\belowdisplayskip{#1}}
{\egroup\ignorespacesafterend}
\begin{shrinkeq5}{-1ex}
\[
w_i = (1 - \alpha) \cdot w_i + \alpha \cdot \sum\nolimits_j f_e(x_{i_j})
\]
\end{shrinkeq5}

where \(\alpha\) is the interpolation coefficient balancing the old weights and the new prototypes. The fine-tuning of \(w\) can be viewed as slightly rotating the classification hyperplane to adapt to shifts in the distribution \(p(x)\), thereby aligning the surrogate with the target distribution. Test-Time Distribution Alignment (TTDA) requires no additional labeled data, with updates performed online during testing. It only fine-tunes the classification head without extra forward or backward passes, resulting in minimal computational overhead.

\section{Experiments}
\subsection{Experimental Settings}
We use ResNet-18 and ResNet-34 \cite{RESNET} trained on CIFAR-10 and CIFAR-100 \cite{cifar} as victims. ResNet-18 is employed as the surrogate architecture to approximate the victims' functionality. We compare $E^3$ against classical and SOTA extraction methods from top-tier AI and security conferences. In the data-dependent (DD) scenario, we include Knockoff Nets \cite{T. Orekondy} from CVPR 2019. In the data-free (DF) scenario, we evaluate techniques including DFME \cite{DFME} from CVPR 2021 and IDEAL \cite{IDEAL} from ICLR 2023, based on deep generative models, as well as CloudLeak \cite{CLOUD} from NDSS 2020, SPSG \cite{Spsg} from CVPR 2024 based on adversarial perturbations, and Marich \cite{P. Karmakar}, the SOTA active sampling strategy from NeurIPS 2023.

For $E^3$, we use the minimal query budget from the comparison methods in both DD and DF scenarios. On CIFAR-10/100, with full resolution $R=32$, we use a reduced resolution of $r=24$ during the first stage of VarRes. For temperature scaling, we set high temperatures of $\tau=10^2$ and $\tau=10^3$ in the DD and DF scenarios, respectively. In the language-guided query selection, we leverage the lightweight text encoder from MobileCLIP \cite{Mobileclip} to generate class name embeddings, using the LSVRC-2012 \cite{IMAGENET} as the candidate OOD set. For test-time distribution alignment, $\alpha$ is set to 1. The code and detailed settings are available at \url{https://github.com/GradOpt/E3}.

\noindent 
\begin{minipage}{0.49\columnwidth}
  \includegraphics[width=\linewidth]{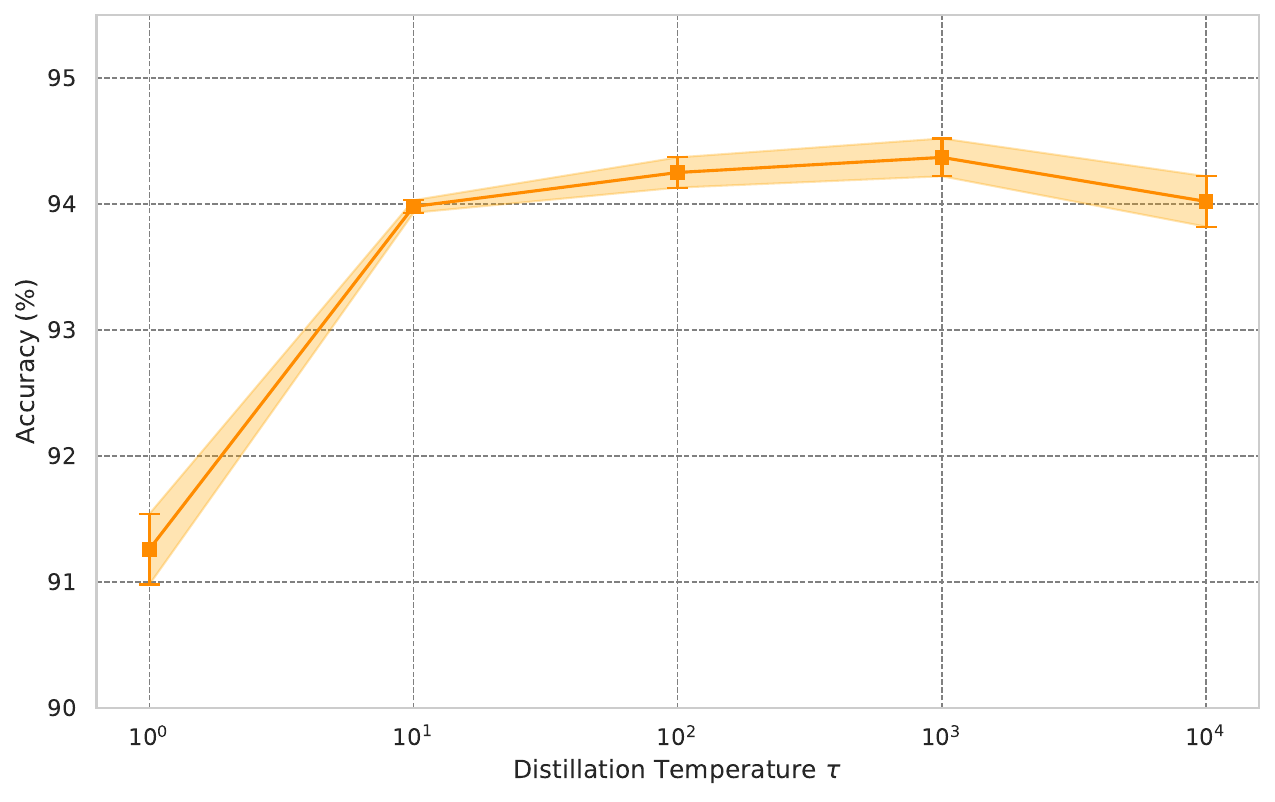} 
  \captionof{figure}{Test Accuracy with Varying Temperatures.} 
  \label{fig:temperature}
\end{minipage}%
\hfill 
\begin{minipage}{0.49\columnwidth}
  \centering
  \includegraphics[width=\linewidth]{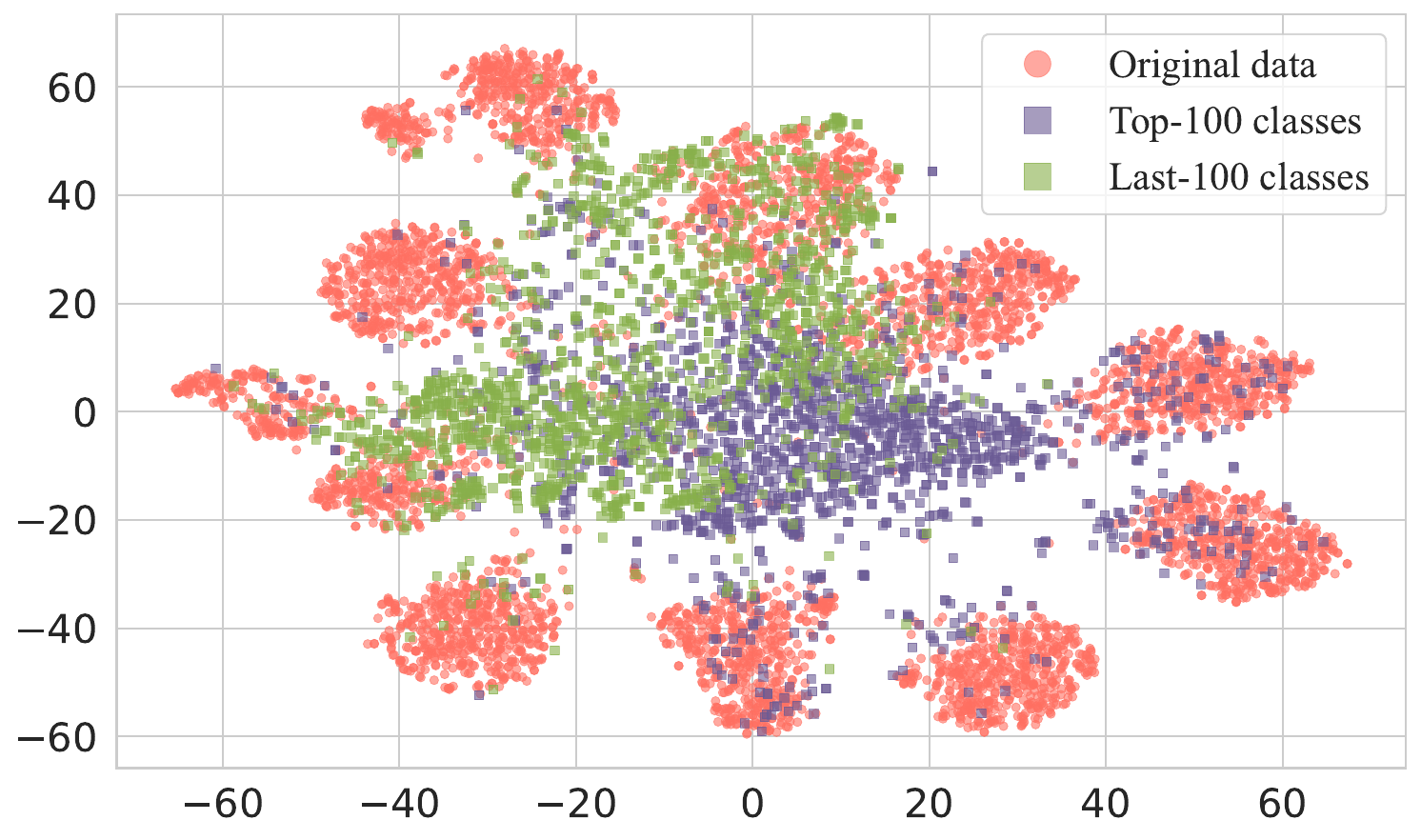} 
  \captionof{figure}{Visualization of Selected Queries in the Feature Space.} 
  \label{fig:feature}
\end{minipage}

\begin{table}[ht]
\centering
\caption{Improvement from TTDA on CIFAR-10 Variants.}
\begin{tabular}{lccc}
\hline
Scenario & Acc w/o TTDA & Acc w/ TTDA & Improvement \\
\hline
CIFAR-10 & 94.35 $\pm$ 0.22 & 94.37 $\pm$ 0.26 & 0.02 \\
CIFAR-10$_{34}$ & 93.94 $\pm$ 0.06 & 94.01 $\pm$ 0.07 & 0.07 \\
CIFAR-10$_{GN}$ & 79.19 $\pm$ 0.85 & 79.71 $\pm$ 0.87 & 0.52 \\
\hline
\end{tabular}
\label{tab:TTDA}
\end{table}

\subsection{Comparison with SOTA Extraction Algorithms}
We compare the query budget, runtime on a single RTX 4090 GPU, and test accuracy of all methods on CIFAR-10/100 in Table \ref{tab:comparison}. $E^3$ consistently outperforms all competitors across different scenarios. VarRes not only reduces runtime but also improves generalization in some cases, allowing the surrogate model to achieve similar functionality to the victim at a lower computational cost. In the DD scenario, $E^3$ requires the same query budget as KnockoffNets but delivers approximately 4\% and 7\% higher accuracy, with a shorter runtime. When both the victim and surrogate use the ResNet-18 architecture, the surrogate trained by $E^3$ even surpasses the victim, benefiting from the regularization effect of knowledge distillation \cite{b21}, while previous extraction methods failed to achieve similar effects due to their inefficient training routines.

In the DF scenario, VarRes exhibits a slight trade-off between performance and efficiency, though generalization remains strong and could be improved by extending the second stage. In contrast, methods based on generative models requires 5 to 10 $\times$ more runtime and a much larger query budget, yet completely failed on more complex tasks like CIFAR-100. Approaches based on adversarial perturbations and active sampling perform slightly better but are still far inferior to $E^3$. Compared to the top-performing competitor, Marich, $E^3$ uses less than half the runtime and achieves over 3\% higher accuracy on both datasets. Notably, $E^3$ in the DF scenario surpasses Knockoff Nets with IND queries, with the surrogate closely approximating the labeling oracle, i.e., the victim.
Interestingly, when the victim becomes more sophisticated (e.g., using ResNet-34 instead of ResNet-18), most surrogates experience a decline in performance, consistent with previous findings \cite{REVISIT}. This can be attributed to the challenge smaller surrogates face in aligning with the victim when the query set is limited. However, $E^3$ exhibits minor degradation, highlighting its potential to effectively target larger and better victims.
\subsection{Ablation Study}
We present a brief analysis of the contributions of each component in $E^3$. Figure \ref{fig:temperature} shows the effect of temperature on CIFAR-10. Higher temperatures significantly improve generalization, and beyond $\tau=10$, accuracy stabilizes, simplifying parameter selection. Figure \ref{fig:feature} visualizes the feature space of ResNet-18, comparing the first and last 100 LSVRC class samples selected via language-guided query selection, along with original CIFAR-10 samples. The first 100 classes effectively capture CIFAR-10 features, demonstrating the efficacy of semantic selection. However, the last 100 classes also provide meaningful features, leading us to sample from all classes with normalized probabilities. This ensures both informativeness and diversity, resulting in a 1.22\% improvement over CIFAR-100 and 0.37\% over random sampling from LSVRC as queries.

Table \ref{tab:TTDA} summarizes the improvements from test-time distribution alignment (TTDA) across various scenarios. Since the surrogate's performance is already close to that of the victim, TTDA provides only a marginal boost. However, when test samples are corrupted by Gaussian noise, TTDA proves particularly effective, making it valuable in cases of distribution shift between the surrogate and victim deployment. Furthermore, TTDA introduces only about 1\% extra latency during deployment and allows model fixation after brief adaptation, resulting in negligible overhead.


\section{Conclusion}
In this paper, we address the inefficiency of current model extraction by thoroughly revisiting design choices throughout the pipeline. From query preparation to training routine and surrogate deployment, we propose $E^3$, a novel algorithm that achieves superior generalization with minimal computational overhead. Our strategies can be seamlessly integrated in a plug and play manner to optimize existing and future extraction algorithms. For future work, we plan to extend $E^3$ to other data modalities and model families \cite{DF}, and to experiment with larger datasets and real-world APIs.
\end{document}